# Digital twin in advanced training of engineering specialists

S N Masaev[1,2,3,6], A N Minkin[1,4], E Yu Troyak[4] and A L Khrulkevich[1,5]

[1]Siberian Federal University, pr. Svobodnyj, 79, Krasnoyarsk, 660041, Russia
[2]Department of System Analysis and Operations Research, Reshetnev Siberian State University of Science and Technology, 31, Krasnoyarsky Rabochy Av., Krasnoyarsk, 660037, Russia
[3]Control Systems LLC, 86 Pavlova Street, Krasnoyarsk, 660122, Russia
[4]FSBEI HE Siberian Fire and Rescue Academy EMERCOM of Russia, 1 Severnaya Street, Zheleznogorsk, 662972, Russia
[5]The Main Directorate of EMERCOM of Russia for Krasnoyarsk Territory, Mira Ave., 68, Krasnoyarsk, 660122, Russia

[6]E-mail: faberi@list.ru

**Abstract**. A review of scientific literature showed the relevance of the issue of assessing the training of an engineering specialist. Engineering includes a variety of works that relate to production issues. To assess the training of an engineering specialist, the digital twin of the enterprise is used. The digital twin of an enterprise includes all major pre-production, production and production support processes. A method for assessing the competencies received by an engineering specialist has been selected. The set of competencies is associated with the ongoing processes of the digital twin of an industrial enterprise. The normal mode of execution of the enterprise production processes is measured. The mode of operation of the processes was measured taking into account the new competencies of an engineering specialist. The study proposes a methodology for assessing the competence of an engineering specialist. The assessment of the effectiveness of the training passed by him.

## 1. Introduction

The knowledge of engineering specialists is implemented in enterprises through competencies. How quickly the employee copes with work tasks depends on the competencies obtained at the university.

Competencies acquired in universities can be represented in different ways. For example, interpret them through Universal Competencies (recommended by the Council of Europe) [1], Dublin Descriptors [2], European Qualifications Framework (passport of qualifications) [3], European Qualifications Framework for EU countries [4], National Qualifications Framework [5]. The above qualification documents are united by the creation of universal competencies within the Bologna process [4]. One of the first works in this area of L A Rastrygin [6].

The process of creating universal competencies generates a huge process of iterating competencies from other methodologies. There is a huge analysis of the applicability of each competency to various industries. After that, update the curricula of educational institutions.

Engineering specialists is no exception. Often engineering specialists of educational institutions do not work in their specialty. All-Russian statistics say that 60% of graduates do not work in their specialty. Consequently, there is a large percentage of competencies not in demand by the employer.

It is urgent to solve the problem of assessing the demand for competencies at enterprises. This will create a list of universal and modern competencies of engineering specialists for the work programs of the discipline at universities. Allows to save resources for retraining of engineering specialists by enterprises.

Therefore, the purpose: assess the relevance of the competencies of engineering specialists at the enterprise after study.

We going to do a tasks for that:

- Simulate enterprise activities (create the digital twin);
- Compare the competencies of engineering specialist with enterprise processes;
- Assess the impact of the competencies of engineering specialist on the activities of the enterprise;
- Analysis of results.

We formalize the enterprise as a dynamic system (model of enterprise) in the classical representation [7, 8].

## 2. Method

Step 1. Many processes of the enterprise are formed $X$. Each $t$ time period is analyzed for the presence of an ongoing or non-running process. Then a system of processes is formed $S=\{T,X\}$, where $T=\{t:t=1,...,T_{max}\}$ - a lot of time points, $x(t)=[x^1(t),x^2(t),...,x^n(t)]^T \in X$ – $n$ – vector of indicators process. Indicators of the vector $x^i(t)$ - the value of financial expenses and income of the enterprise [9,10]. $S=\{T,X\}$ identified as a model of the enterprise. Classical surveillance and control tasks apply to model [7-11].

Step 2. Comparison of competencies of university graduates $v(t)$ with processes $x(t)$ of the enterprise $S=\{T,X\}$. We form $v_i^j$ (compliance $x_j^i$ is $v_i^j$ set as 1-yes, 0-no) from $i$ – from competence of university graduates and $j$ process model. We get competencies of university graduates set $V$ where $v(t) = \left[ v_1^1(t), v_2^j(t), ..., v_m^n(t) \right]^T \in V$.

Payment competencies of the model is limited by resources $C$, then $C(V) \leq C$. This restriction applies to all subsystems of the researched system.

Step 3. Calculation of the integral $V$ index through the correlation matrix $R_i(x)$

$$V_i(t) = R_i(t) = \sum_{j=1}^{n} |r_{ij}(t)|. \tag{1}$$

$$R_k(t) = \frac{1}{k-1} V_k^{oT}(t) V_k^o(t) = \|r_{ij}(t)\|, \tag{2}$$

$$r_{ij}(t) = \frac{1}{k-1} \sum_{l=1}^{k} v^{oi}(t-l) v^{oj}(t-l), \quad i,j = 1,...,n, \tag{3}$$

$$V_k(t) = \begin{bmatrix} v^T(t-1) \\ v^T(t-2) \\ ... \\ v^T(t-k) \end{bmatrix} = \begin{bmatrix} v^1(t-k) & v^2(t-k) & \cdots & v^n(t-k) \\ v^1(t-k) & v^2(t-k) & \cdots & v^n(t-k) \\ ... & ... & ... & ... \\ v^1(t-k) & v^2(t-k) & \cdots & v^n(t-k) \end{bmatrix} \tag{4}$$

where $t$ are the time instants, $r_{ij}(t)$ are the correlation coefficients of the variables $v^i(t)$ и $v^j(t)$ at the time instant $t$.

Step 4. The analysis of experimental data is performed graphically. The dynamics of the integral indicator [9] is calculated for all periods of time.

$$V = \sum_{t=1}^{T=\max} \sum_{i=1}^{n} V_i(t).$$

(5)

## 3. Characteristics of the research objects

The application of competences of engineering specialists graduates is being considered at a woodworking enterprise in the city of Krasnoyarsk, Krasnoyarsk Territory. The average headcount of the enterprise is 650 people.

The competencies of engineering specialists in the specialties of the Siberian Federal University (SFU) are considered.

The competencies of engineering specialists include: understand the product life cycle, the basics of marketing, the basics of mechanical engineering, tracking advanced developments, understand the logistics of production, know the rationing in production, understand the organizational structure of an enterprise, know the range of spare parts, know advanced equipment, understand the technological and design documentation, know the international ISO standards, know the automated management systems, know the customer relationship management system, know the methodology of strategic planning and goal setting, know the theory of risks, know specialized software, know ergonomics, know the terminology in the native and foreign languages, apply mathematical analysis.

Specialties are regulated by the order of the Ministry of Education and Science of the Russian Federation of September 12, 2013 N 1061 "On approval of the lists of specialties and areas of training for higher education" (with amendments and additions).

Competencies in specialties versus enterprise model. The enterprise model is characterized by 1.2 million parameters. Parameters are described in works [11, 12]. Simulation is performed in the software package.

## 4. Experiment result

Initial calculation data: $n=1.2$ million values, $X=5,641,442$ thousand rubles, control is set through competencies of SFU engineering specialists ($V_{SFU}$). From the 7th period, for a permanent job, an engineering specialist is introduced after study in SFU. Also, a business trip of the engineering specialist is paid for training the selected method of control at the enterprise (model). The calculation algorithm is 412 minutes.

The standard operating mode of the enterprise $V_{(basic\ mode)}=5,069.9$ is calculated in the previous work [12].

A table 1 shows the experiment result of estimating the control mode ($V_{SFU}$) through competence of SFU engineering specialists.

Table 1. Regime $V_{(SFU)}$.

| $t$ | $V_{(SFU)}$ | $t$ | $V_{(SFU)}$ | $t$ | $V_{(SFU)}$ | $t$ | $V_{(SFU)}$ |
|---|---|---|---|---|---|---|---|
| 1 | 110.67 | 16 | 70.05 | 31 | 104.10 | 46 | 103.64 |
| 2 | 90.31 | 17 | 101.55 | 32 | 97.66 | 47 | 87.22 |
| 3 | 68.99 | 18 | 132.62 | 33 | 81.19 | 48 | 68.27 |
| 4 | 81.36 | 19 | 132.20 | 34 | 75.23 | 49 | 54.14 |
| 5 | 90.07 | 20 | 153.90 | 35 | 68.52 | 50 | 62.51 |
| 6 | 110.04 | 21 | 164.54 | 36 | 60.52 | 51 | 49.02 |
| 7 | 123.53 | 22 | 151.03 | 37 | 53.13 | 52 | 60.58 |
| 8 | 124.07 | 23 | 144.75 | 38 | 61.65 | 53 | 59.33 |
| 9 | 125.43 | 24 | 119.10 | 39 | 53.51 | 54 | 147.33 |

| 10 | 100.32 | 25 | 91.02  | 40 | 51.84  | 55    | 158.41   |
|----|--------|----|--------|----|--------|-------|----------|
| 11 | 74.76  | 26 | 104.59 | 41 | 72.03  | 56    | 156.87   |
| 12 | 72.97  | 27 | 91.59  | 42 | 93.08  | 57    | 167.90   |
| 13 | 79.69  | 28 | 79.04  | 43 | 99.23  | Total | 5,491.33 |
| 14 | 90.80  | 29 | 76.26  | 44 | 115.80 |       |          |
| 15 | 67.93  | 30 | 95.32  | 45 | 110.12 |       |          |

A figure 1 shows the experiment result of estimating the control mode $V_i(t)$ through competence of SFU engineering specialist.

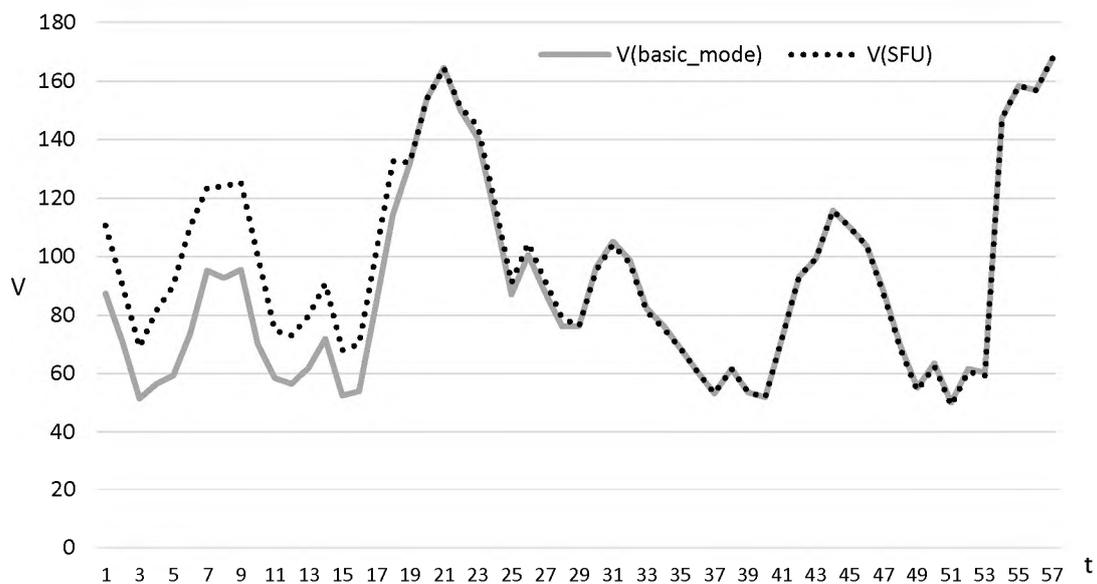

**Figure 1.** Indicator dynamics $V_i(t)$.

## 5. The discussion of the results

Their salaries were increased, additional resources were issued in the amount of 697 thousand rubles. Then the total costs of the enterprise for five years will amount to 5,642,139 thousand rubles. This means that the indicator records organizational measures for the implementation of the assessment of the application of the competencies of SFU engineering specialist at the enterprise.

A unified assessment of the application of the competencies of engineering specialist at the enterprise allows us to determine the most in demand. Sort the most important competencies according to the processes performed in the enterprise. From the sorted competencies to strengthen existing curricula or create new ones at the university. A promising direction is the creation of specialized educational disciplines for the retraining of general directors of enterprises or engineers of very high qualifications. Graduates of special courses must be able to master related fields as well as knowledge of their field.

The approach proposed in the work can be used for other specialties of SFU: Mathematician, Physicist, Programmer, Lawyer, Manager, Marketer, Architect, Systems Analyst, Fire safety specialist, Occupational safety specialist, Social manager, Chemist, Geologist, Mine surveyor, Metal forming specialist, Biochemist, Petrochemistry, Geographer, Bioengineer, Designer, Builder, Real estate appraiser, Automation specialist, Radio technician, Signalman, Nuclearman, Hydroelectric specialist, Mechanic, Constructor, Food technologist, Agronomist, Driller, Metallurgist, Logist, Metrologist, Quality specialist, Nanotechnologist, Psychologist, Economist, Accountant, Commodity specialist, PR man, Journalist, Tourism specialist, Teacher, Philologist, Linguist, Historian, Document flow specialist, Philosopher, Physical Education Specialist, Producer, Culturologist.

The article is an important step for assessing the application of the competencies of graduates of other educational institutions at various enterprises or economic entities.

## 6. Conclusion
The research tasks were completed:

- Modeled activity of the enterprise $S=\{T,X\}$ (digital twin);
- Competencies of engineering specialists were compared with enterprise processes $v(t) = \left[v_1^1(t), v_2^j(t), ..., v_m^n(t)\right]^T \in V$;
- Assess the impact of the competencies of university engineering specialist on the activities of the enterprise $V_{(SFU)}$;
- Results analysis performed $\Delta V = V_{(SFU)} - V_{(basic\_mode)} = 5{,}491.33 - 5{,}069.93 = 421.40$.

The purpose of the research has been achieved.